\documentclass[universe,communication,accept,pdftex,moreauthors,pdftex]{Definitions/mdpi} 
\makeatletter
\let\c@lofdepth\relax
\let\c@lotdepth\relax
\makeatother

\usepackage{subfigure}

 \newcommand\beq{\begin{equation}}
 
 \newcommand\eeq{\end{equation}}
 \newcommand\beqn{\begin{eqnarray}}
 \newcommand\eeqn{\end{eqnarray}}

\def\GeV{\,\mbox{GeV}}

\def\Im{\,{\rm Im}\,}

\Title{Coulomb-Nuclear Interference in Polarized pA Scattering}

\TitleCitation{Coulomb-Nuclear Interference in Polarized pA Scattering}

\Author{Boris  Kopeliovich $^{1,}$*, Michal Krelina $^{2}$ and
Irina Potashnikova $^{1}$}

\AuthorNames{Boris Kopeliovich, Michal Krelina, Irina Potashnikova}
\firstpage{1} 
\pubvolume{1}
\issuenum{1}
\articlenumber{0}
\pubyear{2024}
\copyrightyear{2024}
\datereceived{} 
\daterevised{} 
\dateaccepted{} 
\datepublished{ } 
\hreflink{https://doi.org/}

\AuthorCitation{Kopeliovich, B.; Krelina, M.; Potashnikova, I.}

\address{
$^{1}$ \quad 
Universidad  T\'ecnica  
 Federico Santa Mar\'{\i}a,
Avenida Espa\~na 1680, 2390123 Valpara\'iso,  Chile; 
{irina.potashnikova@usm.cl}\\

$^{2}$ \quad Czech Technical {University} in Prague, Faculty of Nuclear Sciences and Physical Engineering (FNSPE), B\v rehov\'a 7, 11519
Prague, Czech Republic; {michal.krelina@cvut.cz}\\

}
\corres{Correspondence: boris.kopeliovich@usm.cl}

\abstract{
We made the first attempt to understand the observed unusual t dependence of single-spin asymmetry observed in the HJET experiment at RHIC.  
Usually, the interaction of hadrons is presented as a long-range Coulomb interaction and a short-range strong interaction with Coulomb corrections. Such a division gives rise to a Coulomb phase of the hadronic term. Conversely, here we consider short-range hadronic interaction as a correction to the long-range electromagnetic term, i.e., we treat it as an absorptive correction.
This significantly affects the Coulomb-nuclear interference, which is a source of single-spin azimuthal asymmetry at small angles. }


\begin{document}

\section{Introduction}

{Elastic scattering is usually characterized by spin non-flip $f_{nf}$ and spin-flip 
$f_{sf}$  amplitudes, which determine 
the differential elastic cross-sections and single-spin azimuthal asymmetry $A_N(t)$,   }
\beqn 
  \frac{d \sigma_{el}}{dt} &=&   \pi \left(|f_{nf}|^2 + |f_{sf}|^2 \right),
\label{eq:xsec} \\
  A_N \frac{d \sigma_{el}}{dt} &=& -2\pi\,\textrm{Im} \left( f_{nf}f_{sf}^* \right),
   \label{eq:AN} 
\eeqn
where $t$ is 4-momentum transfer squared. The~latter contains interference of the helicity amplitudes, which makes polarization effects sensitive to the hadron interaction \mbox{dynamics \cite{Bilenky:1964pm,buttimore}}.

 According to Equation~(\ref{eq:AN}), an important condition for azimuthal asymmetry is the existence of a phase shift between the spin amplitudes. However, to~the best of our knowledge, the~phases 
of $f_{nf}$ and $f_{sf}$ at high energies are similar. For example,~in the Regge pole model,  $f^h_{sf}$  and $f^h_{nf}$
have the same phase given by the signature factor. To~maximize $A_N(t)$, one should combine hadronic and electromagnetic amplitudes~\cite{kl}. While the former is predominantly imaginary, the~latter is nearly real. Apparently, a~sizable effect is expected at small momentum transfer squared $t$, where the Coulomb and hadronic amplitudes are of the same order. The~$t$ dependence of asymmetry in $pp$ scattering is described in a simplified approximation~\cite{kl} by
\beq
A_N^{pp}(t)=A_N^{pp}(t_p)\,\frac{4y^{3/2}}
{3y^2+1},
\label{eq:AN-pp}
\eeq
where $y=-t/t_p^{pp}$ and
\beqn
t^{pp}_p&=&\frac{8\sqrt{3}\pi\alpha_{em}}{\sigma^{pp}_{tot}};
\label{eq:tp}\\
A_N^{pp}(t_p)&=&
\frac{\sqrt{3t_p}}{4m_p}\,(\mu_p-1).
\label{eq:AN-tp}
\eeqn

Here, $\alpha_{em}=1/137$ is the fine structure constant; $\mu_p=2.79$ is the proton magnetic moment. The~asymmetry $A_N^{pp}(t)$ reaches a maximum (\ref{eq:AN-tp}) of about 4--5\%, at~$t=t_p\approx 2\times 10^{-3}\GeV^2$. For~the sake of simplicity, we assume here, like in~\cite{kl}, a~pure non-flip and imaginary hadronic amplitude, no Coulomb phase, etc. In~what follows, we present most of these simplifications. 

The CNI asymmetry Equation~(\ref{eq:AN-pp}) predicted in~\cite{kl} was confirmed by measurements in~\cite{704,poblaguev}.
{It is worth mentioning at least two important practical applications of the CNI effect.
First, it is a unique opportunity to measure the spin-flip component of the hadronic amplitude at high energies} \cite{kz89,kkp}.

{Second, an important application of CNI-generated azimuthal asymmetry is the possibility to measure the polarization of the beam, in~particular at RHIC. The usage of CNI as a polarimeter has been intensively studied in} \cite{Poblaguev:2022hqh,Poblaguev:2022xoa,AbdulKhalek:2021gbh,Poblaguev:2019saw,Poblaguev:2019qwk,Buttimore:2013rez,Buttimore:2012bha,Buttimore:2011zz,Akchurin:2011zzb,Buttimore:2008zz,Buttimore:2007cj,Buttimore:2007zz,Buttimore:2001df,Bates:2000uw}.

Similar relations can be applied to proton--nucleus elastic scattering, 
\beq
t^{pA}_p=K_A\,t^{pp}_p,
\label{eq:KA}
\eeq
with
\beq
K_A= \frac{Z\sigma_{tot}^{pp}}{\sigma_{tot}^{pA}}.
\label{eq:tAp}
\eeq

Correspondingly, the~maximal value of $A_N$ at $t=t^{pA}_p$ reads,
\beq
A_N^{pA}(t^{pA}_p) = \sqrt{K_A}\, A_N^{pp}(t_p).
\label{eq:AN-Ap}
\eeq

A simple estimate $\sigma_{tot}^{pA}$$\sim$$A^{2/3}\sigma_{tot}^{pp}$ leads to non-dramatic modification of the CNI asymmetry, with about a 20\% increase in $A_N^{pA}(t^{pA}_p)$
even for heavy nuclei, e.g.,~gold.

The energy of $100\GeV$ in the Lab frame is not sufficiently high to suppress iso-vector Reggeons, which have quite a large (dominating) spin-flip component. This is why it is difficult to disentangle  large Reggeon and small Pomeron spin-flip terms. On~iso-scalar nuclei, e.g.,~carbon, copper, etc., iso-vector Reggeons are completely excluded, otherwise they are suppressed by a small factor $(A-2Z)/A$.

\section{Born~Approximation}

The elastic $pA$ amplitude is fully described by two  helicity amplitudes $f_{nf, sf}$ defined in~\cite{buttimore}, each having hadronic and electromagnetic  parts,
$f_i(q_T) = f^h_i(q_T) + f_i^{em}(q_T)$, where for small-angle elastic scattering $t\equiv -q^2\approx -q_T^2$, {they are comparable.}

Hadronic Born amplitudes can be represented as 
\beq 
  \left. f_{nf}^{pA(h)}(q_T)\right|_B =  \frac{\sigma_{tot}^{pA}}{4\pi} F_A^h(q_T^2),
  \label{eq:pA:fHnfB}
  \eeq
  
  \beq
  \left.f_{sf}^{pA(h)}(q_T)\right|_B =  r_5^{pA} \frac{q_T}{m_N}\frac{\sigma_{tot}^{pA}}{4\pi} \textrm{Im} F_A^h(q_T^2),
  \label{eq:pA:fHsfB}
\eeq
where $r_5^{pA}$ is a nuclear analog of  $r_5^{pp}$ defined in~\cite{buttimore}
\beq
r_5^{pp} =
\frac{m_N\, f_{sf}(q_T)}
{q_T\Im f_{nf}(q_T) }.
\label{eq:r5-pp}
\eeq

For $pp$ elastic scattering,  $r_5^{pp}$ was fitted to data in~\cite{kkp,poblaguev}. 

The hadronic nuclear form factor  
$F_A^h(q_T^2)$ in Equations~(\ref{eq:pA:fHnfB})--(\ref{eq:pA:fHsfB}) can be evaluated within the Glauber approximation~\cite{Kopeliovich:2003tz},
\beq 
  F_A^h(q_T)
   = \frac{2i}{\sigma_{tot}^{pA}} \int d^2b \, e^{i \vec q_T \cdot \vec b} \left[ 1 - \left(1 - \frac{1}{2A}\sigma_{tot}^{pp} (1 - i\rho_{pp}(s)) T_A^h(b) \right)^A \right],
\label{eq:ff}
\eeq
as well as the total nuclear cross-section,
\beq
\sigma_{tot}^{pA} = 2  \textrm{Im} \int d^2b \,i \left[ 1 - \left(1 - \frac{1}{2A}\sigma_{tot}^{pp}(s) (1 - i\rho_{pp}(s)) T_A^h(b) \right)^A \right].
\label{eq:sigtot}
\eeq

Correspondingly, the~Born electromagnetic amplitudes read

\beq
   \left. f_{nf}^{pA(em)}(q_T)\right|_B =  -2Z\alpha_{em}\frac{1}{q_T^2 + \lambda^2} F_A^{em}(q_T),
  \label{eq:pA:fEMnfB}
  \eeq
  
  \beq
  \left.f_{sf}^{pA(em)}(q_T)\right|_B= -Z\alpha_{em}\frac{1}{m_N q_T} (\mu_p - 1) F_A^{em}(q_T),
  \label{eq:pA:fEMsfB}
\eeq
where the small fictitious photon mass $\lambda$ is introduced to avoid infrared divergence.
The final results are checked for stability at $\lambda\to0$.

The nuclear electromagnetic form factor in Equations~(\ref{eq:pA:fEMnfB}) and (\ref{eq:pA:fEMsfB}) has the form
\beq
 F_A^{em}(q_T)
= \frac{1}{Z} \int d^2 b e^{i \vec q_T \cdot \vec b} T_Z(b),
\label{eq:ff-em}
\eeq

The nuclear thickness functions are defined as
\beq
    T_A(b) = \int_{-\infty}^{\infty}dz \, \rho(b,z),
    \label{eq:TA}
    \eeq
where $\rho(b,z)$ is the nuclear density distribution function, and~$T_Z(b)=T_A(b)\,Z/A$.

Following~\cite{Kopeliovich:2003tz,Kopeliovich:2005us}, next we  replace the nuclear thickness function with a more accurate effective thickness function convoluted with $NN$ elastic amplitude
\beq
    T_A(b) \Rightarrow  T_A^h(b),    
    \label{eq:TA(h)}
\eeq
where
\begin{equation}
    T_A^h(b) = \frac{2}{\sigma_{tot}^{hN}} \int d^2s\, \textrm{Re} \Gamma^{hN}(s) T_A(\vec b - \vec s);
\end{equation}
\begin{equation}
    \textrm{Re} \Gamma^{hN}(s) = \frac{\sigma_{tot}^{hN}}{4\pi B_{hN}} \exp\left( -\frac{s^2}{2B_{hN}} \right),
\end{equation}
where $B_{hN}$ is the slope of the differential $hN$ elastic cross-section.
This effective nuclear thickness function can be simplified to
\begin{eqnarray}
  T_A^h(b)
   &=& \frac{2}{\sigma_{tot}^{hN}} \int d^2s\, \frac{\sigma_{tot}^{hN}}{4\pi B_{hN}} \exp\left( -\frac{s^2}{2B_{hN}} \right) T_A(\vec b - \vec s) \nonumber\\
   &=& \frac{1}{2\pi B_{hN}} \int d s d\phi\, s \exp\left( -\frac{s^2}{2B_{hN}} \right) T_A(\sqrt{b^2+s^2-2bs\cos\phi}).
\end{eqnarray}

 \section{Hadronic vs. Electromagnetic~Amplitudes}
 
\begin{figure}[H]
  \includegraphics[scale=0.4]{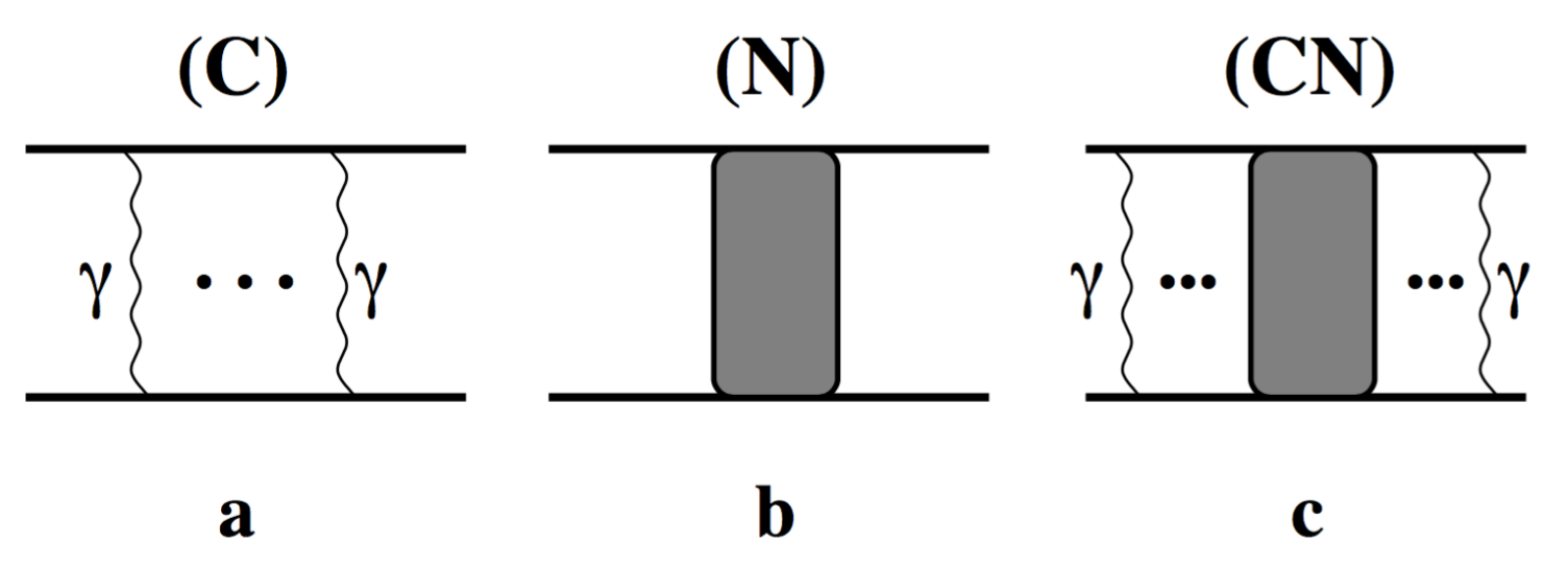}
  \caption{Three types of interaction: pure electromagnetic ({\bf a}),
pure strong interaction ({\bf b}), and~combined strong and electromagnetic interactions ({\bf c}). }
  \label{fig:graphs}
\end{figure}

 The long-range Coulomb forces affect the strong-interaction amplitude by generating a phase shift, known as the Coulomb phase~\cite{bethe,cahn,Kopeliovich:2000ez}. The~interplay of Coulomb and hadronic interaction mechanisms is illustrated in Figure~\ref{fig:graphs}, following the consideration of this problem in~\cite{Kopeliovich:2000ez}.
{In~\cite{Kopeliovich:2000ez} the last two graphs, (N) and (CN) were combined and treated  as a Coulomb-modified strong-interaction amplitude. The~modification was approximated by giving an extra phase factor to the hadronic amplitude. In the literature, this factor is called the Coulomb phase} \cite{bethe,cahn,Kopeliovich:2000ez}.
{Alternatively, one can combine (C) and (CN) and obtain a hadronic correction to the Coulomb amplitude. In~what follows, we call this absorptive {correction}
} \cite{Kopeliovich:1997bp,trueman}.

As usual, multiple interaction amplitudes, depicted in Figure~\ref{fig:graphs}, are easily calculated in impact parameter representation, where the result is just a product of multiple amplitudes. Thus, we switch from $q_T$ to $b$ dependent amplitudes, and~simultaneously, from the Born approximation to the eikonal optical model.
\begin{eqnarray}
\gamma_{nf}^{pA(h)}(b)
    &=&\frac{i}{2\pi}\int d^2q_T\,e^{-i\vec q_T\cdot\vec b}\,
    f_{nf}^{pA(h)}(q_T)
    \label{eq:pA:fEMnf}
\\
\gamma_{sf}^{pA(h)}(b)
    &=&\frac{i}{2\pi}\int d^2q_T\,e^{-i\vec q_T\cdot\vec b}\,
     f_{sf}^{pA(h)}(q_T)
       \label{eq:pA:fEMsf}
       \eeqn
       
These hadronic eikonal phases were used in the Glauber model expressions (\ref{eq:ff}) and (\ref{eq:sigtot}).

Adding higher order terms to the Born Coulomb amplitudes (\ref{eq:pA:fEMnf}) and (\ref{eq:pA:fEMsf}), as~is explained in~\cite{Kopeliovich:1997bp}, we arrive at the eikonal form for the electromagnetic amplitudes
\beq
f_{nf}^{C} (q_T)
  =
    \frac{i}{2\pi} 
    \int d^2b\,e^{i\vec q_T\cdot\vec b}
    \left( 1 - e^{i\chi^C_{nf}(b)}\right),
    \label{eq:eik-nf}
    \eeq
    
    \beq
    f_{sf}^{C} (q_T) 
=
    \frac{1}{2\pi} 
    \int d^2b\,e^{i\vec q_T\cdot\vec b}
    \chi_C^{sf}(b)  e^{i\chi^C_{nf}(b)},
 \label{eq:eik-sf}
    \eeq
 \textls[-25]{with Coulomb eikonal phases, given by the Born amplitudes in impact \mbox{parameter representation.}}

\beqn
\chi^C_{nf}(b) 
    &=& 
    \frac{-2Z\alpha_{em}}{2\pi}\int d^2q_T\,e^{-i\vec q_T\cdot\vec b}\,
    \frac{F_A^{em}(q_T)}{q_T^2 + \lambda^2} 
    \label{eq:chi-nf}
    \\
   \chi^C_{sf}(b) 
    &=&
    \frac{-Z\alpha_{em}(\mu_p-1)}{2\pi m_N}\int d^2q_T\,e^{-i\vec q_T\cdot\vec b}\,
    \frac{F_A^{em}(q_T)}{q_T^2 + \lambda^2} \frac{\vec q_T\cdot\vec b}{b}.
    \label{eq:chi-sf}
\end{eqnarray}

Combining the Coulomb ({\bf {C}}) with Coulomb-nuclear ({\bf {CN}}) mechanisms depicted in Figure~\ref{fig:graphs}, one obtains the Coulomb terms, Equations~(\ref{eq:eik-nf}) or (\ref{eq:eik-sf}), which acquire an absorption factor, given by the standard Glauber eikonal approximation~\cite{glauber},
\beqn
S(b) &=&1-\gamma_{nf}^{pA(h)}(b)=\left(1 - \frac{1}{2A}\sigma_{tot}^{pp}(s) (1 - i\rho_{pp}(s)) T_A^h(b) \right)^A\nonumber\\
&\approx &\exp\left[-{1\over2}\sigma_{tot}^{pp}(s)T_A^h(b)\right].
\label{eq:abs}
\eeqn

As usual, the~complicated multi-loop integrations in momentum representation are essentially simplified to a multiplicative combination in impact parameters. This is why the correction $S(b)$ of Equation~(\ref{eq:abs}) enters the final expressions as a factor.
Thus, the~absorption-corrected amplitudes $\tilde f$ take the form
\beq
\tilde{f}^C_{nf}(q_T)
  =
    \frac{i}{2\pi} 
    \int d^2b\,e^{i\vec q_T\cdot\vec b}
    \left( 1 - e^{i\chi^C_{nf}(b)}\right)
    S(b),
    \label{eq:eik-nfS}
    \eeq
    
    \beq
\tilde{f}^C_{sf}(q_T) 
=
    \frac{1}{2\pi} 
    \int d^2b\,e^{i\vec q_T\cdot\vec b}
    \chi_C^{sf}(b)  e^{i\chi^C_{nf}(b)}S(b),
 \label{eq:eik-sfS}
    \eeq

Eventually, we arrive at the full amplitudes.
\beq
f_{nf,sf}(q_T)=\tilde{f}^C_{nf,sf}(q_T) +f^N_{nf,sf}(q_T).
\label{eq:full-amp}
\eeq

\section{Results vs.~Data}

\begin{figure}[H]
    \includegraphics[scale=0.75]{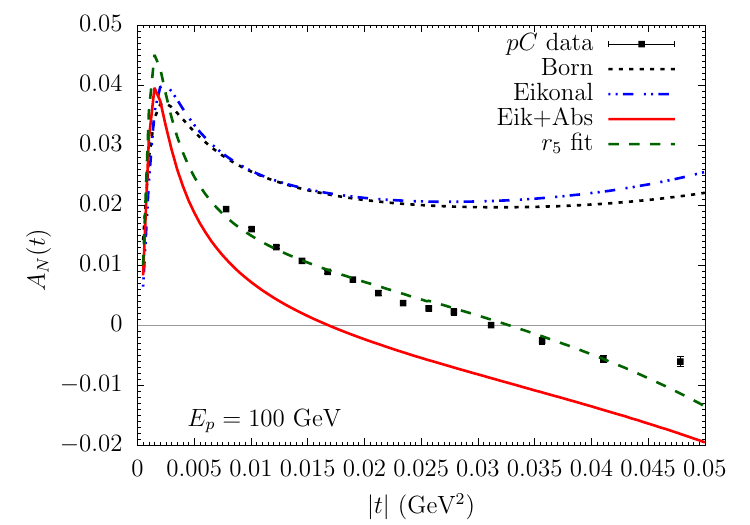}
    \caption{Azimuthal asymmetry in proton--carbon     	 elastic scattering.
    Black dotted curve presents Born approximation, Equations~(\ref{eq:pA:fHnfB})--(\ref{eq:pA:fHsfB}) and (\ref{eq:pA:fEMnfB})--(\ref{eq:pA:fEMsfB}).
    Double-dot-dashed blue and solid red curves correspond to Eikonal approximation without, (\ref{eq:eik-nf})--(\ref{eq:eik-sf}), and~with, (\ref{eq:eik-nfS})--(\ref{eq:eik-sfS}), absorptive corrections, respectively. In~both cases, $r_5^{pA}=0$; data for $pC$ scattering at    $100\GeV$ are from~\cite{Jinnouchi:2004up}.}
    \label{fig:pC}
\end{figure}

\begin{figure}[H]
    \includegraphics[scale=0.75]{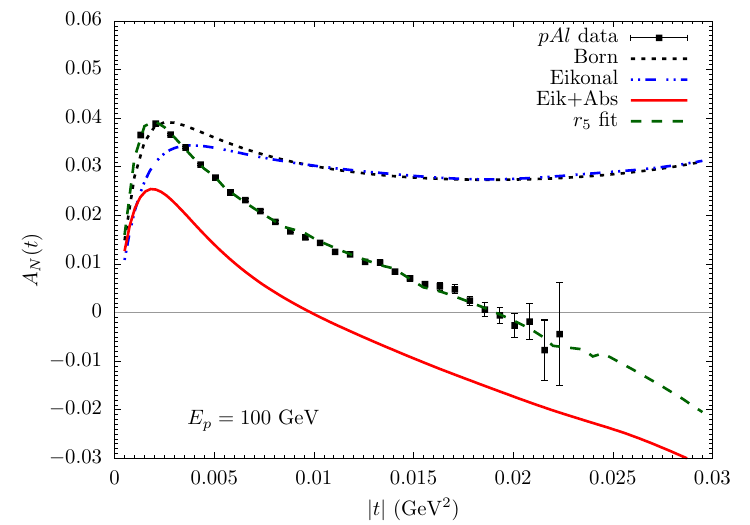}
    \caption{The same as in Figure~\ref{fig:pC}, but~for proton--aluminum elastic scattering. Data at $100\GeV$ are from~\cite{Poblaguev:2015ptq}.}
    \label{fig:pAl}
\end{figure}

\begin{figure}[H]
    \includegraphics[scale=0.75]{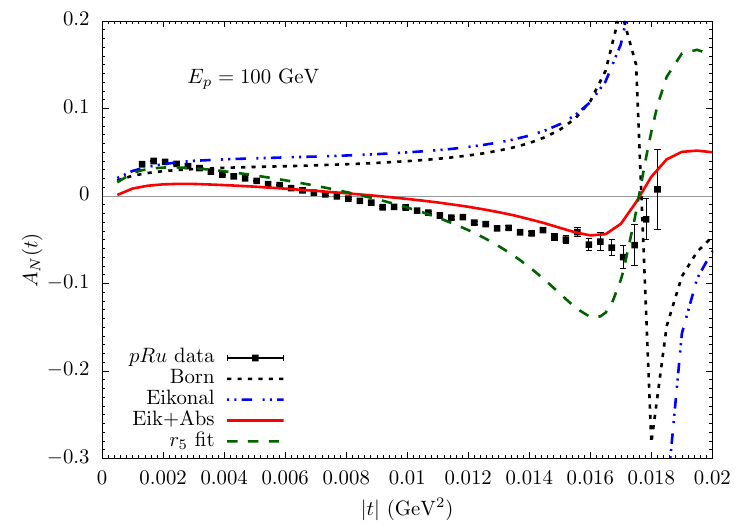}
    \caption{The same as in Figure~\ref{fig:pC}, but~for proton--ruthenium     	 elastic scattering. Data at $100\GeV$ are from~\cite{Poblaguev2023}.}
    \label{fig:pRu}
\end{figure}

\begin{figure}[H]
    \includegraphics[scale=0.75]{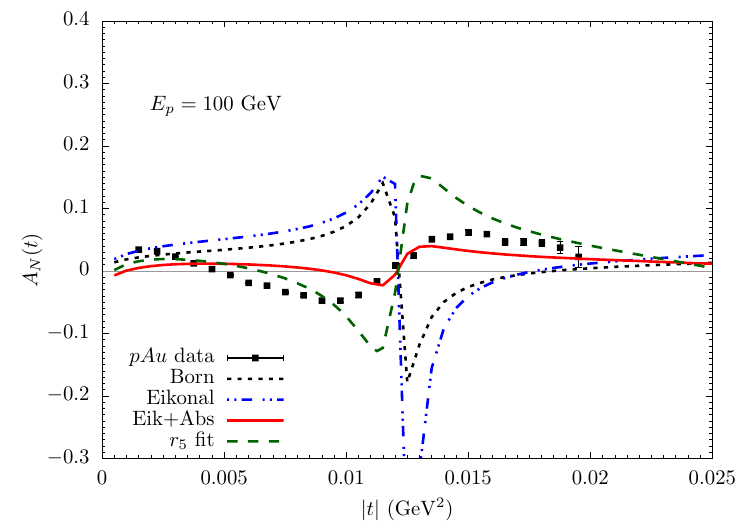}
    \caption{The same as in Figure~\ref{fig:pC}, but~for proton--gold 	 elastic scattering. Data at $100\GeV$ are from~\cite{Poblaguev:2015ptq,Poblaguev2023}.}
    \label{fig:pAu}
\end{figure}

Now we are in a position to calculate the proton--nucleus total cross-section and single-spin asymmetry, given by Equations~(\ref{eq:xsec})--(\ref{eq:AN}). The~results are compared with the data in Figures~\ref{fig:pC}--\ref{fig:pAu}.

\textls[-25]{First of all, we performed calculations within the Born approximation, \mbox{Equations~(\ref{eq:pA:fHnfB}) and (\ref{eq:pA:fHsfB})} and (\ref{eq:pA:fEMnfB}) and (\ref{eq:pA:fEMsfB}). The~hadronic spin-flip component was set to zero. The~results are depicted with black dotted curves in Figures~\ref{fig:pC} and \ref{fig:pAu}. The~magnitude of $A_N$ substantially exceeds the~data.}

Then, we relied on the eikonal form of higher order terms, Equations~(\ref{eq:eik-nf}) and (\ref{eq:eik-sf}), keeping $r_5=0$. The~results are depicted with blue double-dot-dashed curves.
The effect of eikonalization turns out to be rather mild, and the~discrepancy with the data remains~significant.

The next step is the introduction of absorptive corrections, which significantly reduce the values of $A_N(t)$ as is demonstrated by the red solid curves, calculated with Equations~(\ref{eq:eik-nfS}) and (\ref{eq:eik-sfS}) ($r_5$ is still zero). 

Eventually, we can adjust the single unknown parameter, $r_5^{pA}$, for each nuclear target.
The results of the fit are presented in Table~\ref{tab:r5}.

The found values of $r_5^{pA}$ are pretty close to the values for the Pomeron spin component found by fitting to the $pp$ data in~\cite{kkp}. 

With values of $r_5$ in Table~\ref{tab:r5} and amplitude Equations~(\ref{eq:eik-nfS})--(\ref{eq:eik-sfS}), we plot the green dashed curves, which we treat as our final~results.

\unskip

\begin{table}[H]
 \caption{Results of fit for $r_5^{pA}$ to data on $A_N(t)$ for different nuclei~\cite{Jinnouchi:2004up,Poblaguev:2015ptq,Poblaguev2023}.}
    \label{tab:r5}
    \begin{tabularx}{\textwidth}{LCC}
    \toprule
      \textbf{Experiment}   &  \textbf{Re $r_5$ fit}   &  \textbf{Im $r_5$ fit} \\ \midrule
       p-C @ 100~GeV  & $-0.071 \pm 0.003$ & $-0.063 \pm 0.005$\\ 
       p-Al @ 100~GeV  & $-0.080 \pm 0.001$ & $-0.101 \pm 0.002$\\
       p-Ru @ 100~GeV  & $-0.068 \pm 0.001$ & $-0.469 \pm 0.002$\\ 
        p-Au @ 100~GeV  & $-0.027 \pm 0.008$ & $-0.526 \pm 0.006$\\   \bottomrule
    \end{tabularx}
    
\end{table}

Notice that the nuclear data are quite sensitive to the value of $r_5^{pA}$; this is why the CNI method was proposed~\cite{kz89} as a unique way for measuring the hadronic spin-flip amplitudes at high~energies.

\section{Conclusions}

Concluding, we performed the first calculation of single-spin asymmetry in polarized proton--nucleus elastic scattering in the CNI region.
We achieved a reasonable agreement with the data, in~spite of the rather complicated theoretical construction. The~remarkable feature of the nuclear form factor (\ref{eq:ff}) is a change in the sign of the imaginary part of the elastic hadronic non-flip amplitude corresponding to the first zero of the Bessel function $J_1(t)$. Since the non-flip hadronic amplitude significantly exceeds the spin-flip part, they become of the same order right before and after the former changes sign. The~asymmetry $A_N(t)$ Equation~(\ref{eq:AN}) reaches maximum at $f_{nf}=f_{sf}$, so $A_N(t)$ should abruptly vary between positive and negative maxima in the vicinity of the Bessel zero.
Comparison with the data in Figures~\ref{fig:pRu} and \ref{fig:pAu} confirms such behavior; indeed, the measured $A_N(t)$ exposes two maxima with opposite signs with positions close to~predicted. 

However, the magnitude of these maxima is exaggerated in our calculations, so there is still room for improvements. Here, nuclear effects were evaluated within the Glauber approximation, which is subject to Gribov inelastic shadowing corrections~\cite{gribov,gribov-my}. Their calculation requires substantial modeling, including knowledge of the proton wave function, interaction mechanism, etc. This needs a detailed study, as was performed in~\cite{Kopeliovich:2005us}. We leave this issue for future~development.
\vspace{6pt}

\funding{{This work was supported in part by grants ANID---Chile FONDECYT 1231062,~ ANID PIA/APOYO AFB220004,  and~ANID---Millennium Science  Initiative Program 
ICN2019\_044.}}

\begin{adjustwidth}{-\extralength}{0cm}

\reftitle{References}

\PublishersNote{}
\end{adjustwidth}
\end{document}